\documentstyle[11pt]{article}
\input{psfig}
\begin{document}
\begin{titlepage}
\begin{flushright}
{OUTP-99-21-P}\\
\end{flushright}
\vskip 0.5 cm
\begin{center}
 {\Large{\bf  Prediction for $\alpha_3(M_z)$ in a string-inspired model}}\\
        \vskip 0.6 cm
        {\large{\bf Dumitru Ghilencea\footnote{
                E-mail address: D.Ghilencea1@physics.oxford.ac.uk}
}}\\
        \vskip 1 cm
{{\it Department of Physics, Theoretical Physics, University of Oxford}}\\
{\it 1 Keble Road, Oxford OX1 3NP, United Kingdom}\\
\vskip 0.2 cm

\end{center}
\vskip 2 cm
\begin{abstract}
We apply the Renormalisation Group Evolution (RGE) to 
analyse the phenomenological implications of an
extended supersymmetric model, for the value of the 
unification scale  and  the strong coupling at the electroweak scale. 
The model we consider is predicted to exist in 
Calabi-Yau string compactifications with Wilson line mechanism for $E_6$
 symmetry breaking, contains additional matter beyond the 
MSSM spectrum  and avoids the ``doublet-triplet'' splitting problem 
in the Higgs sector. The calculation is analytical in two-loop order
and includes the effects of the heavy thresholds due to the additional
matter considered. The value of $\alpha_3(M_z)$ can be  brought within the 
experimental limits without a significant change of the 
unification scale from  the MSSM prediction. 
\end{abstract}
\end{titlepage}

\setcounter{footnote}{0}
\section{Introduction}
The Minimal Supersymmetric Standard Model (MSSM) currently provides
the ``standard'' framework for the study of the physics beyond the Standard
Model. The MSSM model takes into account the constraints from 
the negative searches so far
for an experimental signature for Supersymmetry and 
gives circumstantial evidence for supersymmetric 
unified theories,  such as the unification of the gauge couplings or
the weak mixing angle prediction. 
The aforementioned constraints and predictions may also be 
fulfilled by other models, of string origin, and this would suggest that
the MSSM is only a {\it minimal} candidate model for the physics 
beyond the Standard Model. Such string inspired models are regarded
as ``low-energy'' limits of string theory, and they 
predict below the compactification
scale  a much richer spectrum (than that of the MSSM) including
additional states  at the high scale corresponding to predicted 
vector-like states\footnote{Vector-like states are not protected by
the chiral symmetry of the SM and are therefore heavy \cite{georgi}.}
under standard model gauge group.
The presence of such additional 
states in conjunction with the low-energy input can 
lead to a phenomenology different from that
of the MSSM. As a specific example which we address in this letter,
consider the measurement of the strong coupling at 
the electroweak scale. The world average value \cite{particledata}
of $\alpha_3(M_z)=0.119\pm 0.002$, situated rather close to the one
loop value predicted in the MSSM is {\it below} the two-loop  MSSM
``bottom-up'' prediction (of $\approx 0.125$) obtained from the 
RGE equations for the gauge couplings.
While at  one-loop level this prediction depends on the 
symmetry of the model and the multiplet content, at two-loop level a
dependence of $\alpha_3(M_z)$ on the high scale thresholds is induced,
and this  may indicate that the mismatch between the MSSM two-loop 
prediction and the experiment is due to our lack of understanding 
of the physics at the high scale. For this reason at least, exploring
the phenomenological implications of string theory {\it predicted}
models with additional intermediate scales finds enough
justification and it may also help in selecting and restricting 
the number of viable string inspired models.

The exact structure of the spectrum that string-inspired models predict to 
exist (in addition to the MSSM spectrum\footnote{
In the following by MSSM spectrum we understand the three generations
of quarks and leptons with their superpartners, an appropriate 
 gauge sector containing the gauge bosons and gauginos and
the two Higgs doublets and superpartners, {\it without} 
any $SU(3)$ Higgs triplets.}) depends in general on the
particular class of  models one considers to investigate.
In reference  \cite{witold1} the authors presented a  
class of extended supersymmetric models, as  the low-energy limit
of a string model with  Calabi-Yau compactification \cite{116} and
Wilson line breaking \cite{wilson} mechanism for the $E_6$ symmetry.

In a  generic example of this class of models, the ``low-energy'' spectrum
below the compactification scale  contained \cite{witold1} the
MSSM spectrum plus (pairs of) complete five and ten-dimensional $SU(5)$ 
multiplets, ``vector-like'' under the Standard Model gauge group.
The phenomenological implications of this case were discussed in 
\cite{grl,grlyuk} for the case when perturbation theory applies up to
the unification scale.
It was found that, due to a ``mixing''
between the heavy thresholds and the two-loop contributions of the
vector-like states to the running of the gauge couplings, 
there is only a small (two-loop) increase in the unification scale 
from the MSSM prediction. The  increase  factor was  $\approx 3$,
too small to make an agreement with the weakly coupled heterotic string
prediction \cite{scale}  which gives an unification scale 
$\approx 20$ times larger than that of the MSSM. 
The aforementioned factor of increase was 
accompanied by a small (two-loop) 
increase from the MSSM prediction for the 
strong coupling at electroweak 
scale\footnote{For the case when the  unification of the 
gauge couplings takes place 
in the non-perturbative regime, when the unified coupling is assumed
to be large, one  way to make predictions was presented 
in \cite{sun}, although the errors which affect them might
be large as discussed in \cite{grlyuk} (Section 5).}.

Reference \cite{witold1} also predicted another interesting 
possibility for the ``low-energy''  spectrum predicted by the same class
of string-inspired  models and this will be 
further analysed in this paper. This more specific model
predicts not only complete five dimensional representations
of $SU(5)$ in addition to the MSSM spectrum, 
but also a ``split multiplet'' structure, triplet under
the $SU(3)$ group of the Standard Model.
The model has therefore \cite{witold1} 
the nice feature of avoiding the ``doublet-triplet''
splitting problem which appears in the context of (Supersymmetric) 
Grand Unified Group Theories (GUT).
In these theories, the Higgs multiplet content is a $5+{\overline 5}$ 
pair and consequently, the bare masses  of the Higgs doublet and Higgs
colour triplet have to be equal. To avoid proton decay mediated by the
latter at a rate forbidden by the  experimental constraints, the Higgs
(colour) triplet has to be heavy enough to suppress such processes. 
In the meantime, the $SU(2)$ Higgs doublet must be light enough  
to explain the mass origin at the electroweak scale. 
This leads to the so-called ``doublet-triplet'' splitting 
problem, specific to  supersymmetric GUT theories as well as 
to other theories which have such Higgs spectrum assignment. 

In addition to avoiding the ``doublet-triplet'' splitting problem, 
this last specific model provides \cite{witold1} the unification 
of the gauge couplings, even in the absence
of a grand unified group such as $SU(5)$ or larger. 
The spectrum predicted by this string inspired model
contains below the compactification scale\footnote{We refer here to
the model which in \cite{witold1} was called  the ``unconventional'' case.}
a pair $3+{\overline 3}$ and an arbitrary number (say $n+1$) of
extra pairs $5+{\overline 5}$ of $SU(5)$ {\it  in addition} to the 
matter fields (three families) and the gauge sector of the 
Minimal Supersymmetric Standard Model, but without its Higgs content
\footnote{Equivalently, we can say that the spectrum just below the
compactification scale (and before the decoupling of any heavy state
as we lower the scale) contains $n$ pairs 
$5+{\overline 5}$ and 2 pairs $3+{\overline 3}$ in addition 
to the MSSM spectrum.}.

A coupling of the  triplet of the 
``incomplete'' $SU(5)$ representation, $3+{\overline 3}$,
 to the triplet component of a  five dimensional representation
of the form  $\lambda {\tilde\phi} 3 {\bar 5}$
can naturally lead to a {\it large} mass (due to their
vector-like character under SM group) term for the triplets, via 
a symmetry breaking mechanism when the (Standard Model singlet)
Higgs field $\tilde\phi$ acquires a v.e.v., through a 
mechanism detailed in \cite{grl}.
The same mechanism applies to 
the {\it extra} $n$ (vector-like) pairs of $5+{\overline 5}$ which also acquire
a {\it large} mass (through couplings $\lambda {\tilde\phi} 5 {\bar
5}$ via the same mechanism \cite{grl} which provides a natural explanation
for the origin of such mass terms). 
The {\it bare} value of this mass, assumed to be the
same for all $5$'s is denoted by $\mu_g$ for further reference.
The (remaining)  doublet components of  
the initial pair $5+{\overline 5}$ are left uncoupled 
and thus light and can therefore account for the Higgs content of the MSSM.

It is the purpose of this letter to examine in 
some detail the phenomenological implications of
this model, in a  two-loop analytical approach. 
The predictions we make refer to the value of $\alpha_3(M_z)$,
the unification scale itself and the scale of the intermediate 
matter\footnote{We take as an input parameter the unified gauge
coupling, to predict the intermediate scale and not vice-versa for
the reason that the intermediate scale tends to have a flat
behaviour for large range of values for $\alpha_g$, which can
induce numerical instabilities of the solution, see
Figures 3 and 4 of ref. \cite{grl} for a similar case.}, $\mu_g$.

The reason for performing a two-loop analytical investigation of
this model is three-fold; as mentioned, the discrepancy 
between the MSSM  prediction for $\alpha_3(M_z)$ and the experiment
arises  mainly due to the two-loop corrections, 
and thus threshold dependence at the high
scale plays a significant role. Moreover, any better candidate model 
than the MSSM should eliminate such discrepancy at this
level of accuracy. Finally, the spectrum predicted by our
model contains in addition to the $SU(3)$ triplet components, complete $SU(5)$
representations which are known to change the low energy prediction
for $\alpha_3(M_z)$ at two-loop level \cite{grl} only\footnote{A similar
situation exists for the unification scale prediction as well.}.
The present  approach also provides an analytical  method to examine,
in two loop order the RGE prediction for the strong coupling and may
be applied to other (string-inspired) models with spectrum different
from that of the MSSM. 

We show that $\alpha_3(M_z)$ can be reduced from the MSSM
value (of $\approx 0.125$ or larger\footnote{The 
exact value of the strong coupling prediction in the MSSM
depends on the assumptions made for the low energy (TeV scale) supersymmetric
spectrum.}) and be brought within the experimental 
limits \cite{particledata} of $0.119\pm 0.002$, while keeping the 
unification scale $M_g$ close to that of the MSSM
for an intermediate scale $\mu_g$ within one order of magnitude below
$M_g$.

\section{Predictions from the Renormalisation Group Evolution}
The standard tool to exploring the phenomenological consequences
of our model is the Renormalisation Group Evolution (RGE)
for the gauge couplings, for which we take as (low-energy) 
boundary conditions
the well known values of $\alpha_1(M_z)$ and $\alpha_2(M_z)$,
obtained from measured  electromagnetic coupling and weak mixing angle
at the electroweak scale. To apply this tool we need to know the 
multiplet content, which was detailed in the Introduction, 
and the symmetry group,
which is {\it just} the Standard Model gauge group, with 
$SU(5)$ normalisation for the $U(1)_Y$ coupling.
Thus, the one-loop beta function before the decoupling of any 
extra (complete or incomplete) multiplet  is  given by
\begin{equation}
b^*_i=b_i+\Delta b_i+n
\end{equation}
with $b_i$   the MSSM one-loop beta function, $b_i=(33/5,1,-3)$, 
and with $\Delta b_i=4 \times\{1/5, 0, 1/2\}$ to  account for two pairs
of triplets or equivalently four triplet  states, hence the factor 4 in 
the definition of $\Delta b_i$. After the decoupling of all additional
states, the one-loop beta function is just that of the MSSM, namely $b_i$.
With some loss of generality we restrict ourselves to the case
when the extra states (triplets and 5-plets) have the same {\it bare} mass
which we called $\mu_g$ and this assumption does not reintroduce
the ``doublet-triplet'' problem. The general case of 
considering different masses for $3+\bar 3$
and $5+\bar 5$ pairs can be done following the present approach, although
introducing one further mass parameter would make the analysis less 
tractable. 

To evaluate the full two-loop ``running'' of the gauge couplings,
including the effects of the heavy thresholds, we use
the integral form of the ``NSVZ beta function''. This has 
been  computed in \cite{novikov} and \cite{murayama1} (see also
\cite{shifman2}) and is given by
\begin{equation}
\beta (\alpha )^{NSVZ}\equiv \frac{d\alpha }{d(\ln \mu )}=-\frac{\alpha ^{2}%
}{2\pi }\left[ 3T(G)-\sum_{\sigma }^{{}}T(R_{\sigma })(1-\gamma _{\sigma
}^{NSVZ})\right] \left( {1-T(G)\frac{\alpha }{2\pi }}\right) ^{-1}
\label{shifmanbetaalpha}
\end{equation}
with the definition ($\mu $ is the running scale) 
\[
\gamma _{\sigma }^{NSVZ}=-\frac{d\ln Z_{\sigma }}{d\ln \mu }
\]
and where $T(G)$ and $T(R_{\sigma })$ represent the Dynkin index for the
adjoint representation and for $R_{\sigma }$ representation respectively
(not necessarily the fundamental one).
The above sum runs over {\it all} matter fields $\sigma $ in
representation $R_{\sigma }$ and this includes the extra heavy
states in addition to the low
energy spectrum of the MSSM. 

Following the details given in \cite{grlyuk} to 
integrate the beta function given above, we find that, to all orders in 
perturbation theory, the gauge couplings run, in the presence of the
extra matter, as follows
\begin{eqnarray} \label{intalph}
\alpha _{i}^{-1}(M_{z}) &=&-\delta _{i}+\alpha _{g}^{-1}+\frac{b_{i}}{2\pi } 
\ln \frac{M_{g}}{M_{z}}+\frac{n+\Delta b_i}
{2\pi }\ln \frac{M_{g}}{\mu _{g}}-\frac{
\beta _{i,H_{1}}}{2\pi }\ln Z_{H_{1}}(M_{z})-\frac{\beta _{i,H_{2}}}{2\pi }
\ln Z_{H_{2}}(M_{z})  \nonumber \\ 
&&-\frac{\beta _{i,g}}{2\pi }\ln \left[ \frac{\alpha _{g}}{\alpha _{i}(M_{z})%
}\right] ^{1/3}-\sum_{j=1}^{3}\sum_{\phi _{j}}{}\frac{\beta _{i,\phi _{j}}}{%
2\pi }\ln Z_{\phi _{j}}(M_{z})
\end{eqnarray} 
where $b_{1}=33/5$, $\,b_{2}=1$, $\,b_{3}=-3$ and where $\beta _{i,\phi
_{j}}\equiv T(R_{\phi _{j}}^{i})$, $i=\{1,2,3\}$, are the contributions to
one-loop beta function\footnote{%
We also used that the one-loop beta function is $b=-3 T(G)+ \sum T(R_\phi)$,
where the sum runs over all chiral supermultiplets in representation $R_\phi$%
.} of the matter fields $\phi _{j}$ (j=generation index), while $\beta
_{i,g}\equiv T^{i}(G)$ is the one-loop beta function for the pure gauge
(+gaugino) sector; the Higgs (+higgsino) sector contribution is included
separately via the terms proportional to $\beta _{i,H1,2}$;
finally, $\alpha_g$ is the unified coupling while $M_g$ stands for the
unification scale of our model. 
We have
\begin{equation}
\beta _{i,\phi _{j}}=\left( 
\begin{array}{ccccc}
\frac{3}{10} & \frac{1}{10} & \frac{3}{5} & \frac{4}{5} & \frac{1}{5} \\ 
&  &  &  &  \\ 
\frac{1}{2} & \frac{3}{2} & 0 & 0 & 0 \\ 
&  &  &  &  \\ 
\ 0 & 1 & 0 & \frac{1}{2} & \frac{1}{2}
\end{array}
\right) _{i,\phi _{j}}\;\;\;\;\;\;\;\beta _{i,g}=\left( 
\begin{array}{c}
0 \\ 
\\ 
-6 \\ 
\\ 
-9
\end{array}
\right) ;\;\;\;\;\;\;\;\beta _{i,H_{1,2}}=\left( 
\begin{array}{c}
\frac{3}{10} \\ 
\\ 
\frac{1}{2} \\ 
\\ 
0
\end{array}
\right)
\end{equation}
with $\beta _{i,\phi _{j}}$ 
independent of the values of $j$. The field $\phi _{j}$ runs over the set $%
\phi _{j}=\{l_{L},q_{L},e_{R},u_{R},d_{R}\}_{j}$, in this order, with $j$ as
generation index. The coefficients $\delta_i$ in eq.(\ref{intalph}) 
represent the low energy supersymmetric thresholds and they would be equal
to zero if supersymmetry were
 valid at the electroweak scale. Their exact expressions
will not be of concern to us as we will present our phenomenological
predictions as a change to the MSSM predictions\footnote{
The MSSM quantities used as an input in our 
calculation will be the unified coupling, the unification scale and
the value of the strong coupling at electroweak scale.}
 which implicitly contain
the dependence on $\delta_i$. ($\delta_i$ also contain conversion
scheme factors (${\overline {MS}}\rightarrow {\overline {DR}}$)
which in our calculation will  cancel against those of the MSSM
of equal value). The uncertainty in the low-energy 
supersymmetric spectrum (i.e. the value of $\delta_i$)
can  be taken into account in our present approach
by allowing in our final results,
a {\it range} of values for the MSSM variables which are present
in our final expressions.

To compute the two-loop running for the gauge couplings, 
only a one-loop expression for the wavefunction renormalisation 
coefficients is required.
Note that in the two-loop approximation there is no regularisation
ambiguity which arises only in  three-loop order \cite{jones}.
At  $M_z$ scale the one-loop expressions for $Z$'s
 have the following structure
\begin{eqnarray} \label{zzss}
Z_{F}(M_z)&=&\prod_{k=1}^{3} \left[\frac{\alpha_g}{\alpha_k(\mu_g)}
\right]^{-\frac{2 C_k(F)}{b^*_k}} 
\left[\frac{\alpha_k(\mu_g)}{\alpha_k(M_z)}\right]^{-\frac{2
C_k(F)}{b_k}}
\nonumber\\
&=&\prod_{k=1}^{3} \left[\frac{\alpha_g}{\alpha_k(\mu_g)}
\right]^{\frac{2 C_k(F)}{b_k}\frac{n+\Delta b_k}{b^*_k}} 
\left[\frac{\alpha_g}{\alpha_k(M_z)}\right]^{-\frac{2 C_k(F)}{b_k}}
\end{eqnarray} 
where $F$ stands for any Higgs or MSSM  chiral field.
Strictly speaking in the expressions of $Z$ factors
 we should have used the mean mass of the extra
states ${\tilde \mu}$ instead of $\mu_g$; however 
this difference is an additional radiative effect and thus 
is of two-loop order for $Z$'s or of three loop order for
the gauge couplings, and can be neglected in our two
loop calculation.
From equations (\ref{intalph}) and (\ref{zzss}) we find 
the following RGE equations 
\begin{eqnarray}  \label{HHMSSM}
&\alpha_i^{-1}(M_z)=&-\delta_i+\alpha_g^{-1}+
\frac{b_i}{2\pi}\ln \left[\frac{%
M_g}{M_z}\right]+\frac{n+\Delta b_i}{2\pi}
\ln\left[\frac{M_g}{\mu_g} \right]-\frac{1}{%
2\pi}\sum_{j=1}^{3}{\tilde Y}_{ij}\ln
\left[\frac {\alpha_g}{\alpha_j(\mu_g)}\right] 
\nonumber \\
& &+\frac{1}{4\pi}\sum_{j=1}^{3} \frac{b_{ij}}{b_j}
\ln\left[\frac{\alpha_g}{%
\alpha_j(M_z)}\right]
\end{eqnarray}
where 
\begin{equation}  \label{whywhy}
{\tilde Y}_{ij}=\frac{n+\Delta b_j}{b^*_j}
\left[\frac{1}{2}\frac{b_{ij}}{b_j}-\delta_{ij}\lambda_j\right]
\end{equation}
with $\lambda_1=0,\,\lambda_2=2, \,\lambda_3=3$ while $\delta_i$,
 $i=\{1,2,3\}$, stand for the low-energy (TeV scale) 
supersymmetric thresholds. 

From  eq.(\ref{HHMSSM}) we can again see 
that the presence of $\mu_g$ in the two-loop 
 term  $\ln(\alpha_g/\alpha(\mu_g))$
instead of the mean physical mass ${\tilde\mu}$ of the additional 
multiplets we consider would account for an additional
(three-loop) radiative effect and we neglect it,
as it is beyond our two-loop approximation for the running of 
the gauge couplings.
Thus, we can say that in two-loop order
the extra states contribute to the gauge couplings running
through their common bare mass only.

At this point we would like to emphasize that the result
of equation (\ref{HHMSSM}) can also be obtained
using the ``standard'' RGE equations, integrated in two-loop order, 
with appropriate taking into account of the heavy thresholds that 
the additional states we consider bring in.
Using a radiative dressing of the masses of the additional states
and following the approach of ref. \cite{grl}
one will obtain the same result. As it was the case
there, there is a  cancellation of 
the heavy thresholds against the two-loop contributions
of the additional states we considered and as a  consequence 
in eq.(\ref{HHMSSM}) only the two-loop MSSM beta function appears, 
and {\it not} that
in the presence of additional states\footnote{Also note the strong similarity
of eq.(\ref{whywhy}) with that of eq.(7) of ref.\cite{grl}.} 
$5+{\overline 5}$ or $3+{\overline 3}$.
Eq.(\ref{HHMSSM}) is actually more general; if one considers vector-like 
matter in addition to the MSSM sector, with some arbitrary  $\delta b_j$  
contribution to  one-loop beta function,  the  two-loop RGE equations have
a form similar to that of eqs.(\ref{HHMSSM}) and (\ref{whywhy}) with the
replacements  $n+\Delta b_i\rightarrow \delta b_i$ 
and $b_j^*\rightarrow b_j+\delta b_j$.

To compute the unification scale $M_g$, the strong coupling
$\alpha_3(M_z)$ and the value 
of the mass scale $\mu_g$ we must make some assumptions about the low
energy supersymmetric spectrum which affects the running of the gauge
couplings through the terms $\delta_i$,  as seen from eq.(\ref{HHMSSM}).
Since the effects of the low energy supersymmetric 
thresholds on the predictions
of  the MSSM  are relatively known \cite{lan},
we prefer to express our predictions as a change to the
MSSM predictions which all have this dependence included
(and assume that $\delta_i$'s have equal values to those of the MSSM). 
We therefore consider the two-loop running of the gauge
couplings in the MSSM, which is of the form
(the MSSM variables are labelled with an ``o''
index to distinguish them from those of our extended model)
\begin{equation}
\alpha _{i}^{o-1}(M_{z})=-\delta _{i}+\alpha _{g}^{o-1}+\frac{b_{i}}{%
2\pi }\ln \left[ \frac{M_{g}^{o}}{M_{Z}}\right] +\frac{1}{4\pi }%
\sum_{j=1}^{3}\frac{b_{ij}}{b_{j}}\ln \left[ \frac{\alpha _{g}^{o}}{\alpha
_{j}^{o}(M_{Z})}\right]  \label{MSSM}
\end{equation}
We  can then  substitute\footnote{In the MSSM
we have $\alpha_g^o\approx 0.0433$, $M_g^o\approx 3\times 10^{16}$ GeV
and $\alpha_3^o(M_z)=0.125$ or larger. These results are obtained in two
loop approximation, using $\alpha_1(M_z)$ and $\alpha_2(M_z)$
as an input from experimental values for electromagnetic coupling and
weak mixing angle.}
 the values of $\delta_i$ from the above
equation into eq.(\ref{HHMSSM}) and impose that
in both the MSSM and our model, the values of $\alpha_1(M_z)$ 
and $\alpha_2(M_z)$ are taken equal with the corresponding experimental
value, so $\alpha_1^o(M_z)=\alpha_1(M_z)$ and
$\alpha_2^o(M_z)=\alpha_2(M_z)$.  We then 
compute the  analytical expressions for
the factor of increase of the unification 
scale $M_g/M^o_g$, the strong coupling $\alpha_3(M_z)$ (in terms of
$\alpha_3^o(M_z)$), and  the bare mass of the extra states, $\mu_g$. 
This can be done following
the approach of \cite{grl} and using that, under two-loop terms 
we can substitute the arguments of the ``log'' terms by their one 
loop values, as the difference would be of higher order. 
This means that $\ln(\alpha_g/\alpha_j(\mu_g))=\ln(1+b_j^*\alpha_g/(2\pi)
\ln(M_g/\mu_g))$ and we further replace $\ln(M_g/\mu_g)$
by its one loop analytical expression, which is correct for a two loop
running for the gauge couplings.

After some tedious algebra we find the following two-loop analytical results
\begin{eqnarray}  
\label{ores1}
\frac{M_g}{M_g^o}&=&
exp\left\{\frac{2\pi}{7n-1}\left(\alpha_g^{-1}-\alpha_g^{o-1}\right)\right\}
\left[\frac{\alpha_g}{\alpha_g^o} \right]^{{\cal E}_1}
\left\{1+\frac{18\,\alpha_3^o(M_z)}{1-7n}
\left(\alpha_g^{-1}-\alpha_g^{o-1}\right)\right\}^{{\cal E}_2}
\nonumber\\
&&\times\prod_{j=1}^{3} \left\{1+\frac{7b^*_j}{1-7n}
\left[1-\frac{\alpha_g}{\alpha_g^o}\right]\right\}^{{\cal D}_j}
\end{eqnarray}
\begin{eqnarray}  
\label{ores4}
\alpha_3^{-1} (M_z)&=&\alpha_3^{o-1}(M_z)+
\frac{18(\alpha_g^{-1}-\alpha_g^{o-1})}{1-7n}
+\frac{73+17n}{2(7n-1)\pi}\ln\left\{1+\frac{18\,\alpha_3^o(M_z)}{1-7n}
\left(\alpha_g^{-1}-\alpha_g^{o-1}\right)\right\}\nonumber\\
&&+\frac{2(467+235n)}{11\pi(1-7n)}\ln\left[\frac{\alpha_g}{\alpha_g^o}\right]
+ \sum_{j=1}^{3} \ln\left\{1+\frac{7b^*_j}{1-7n}
\left[1-\frac{\alpha_g}{\alpha_g^o}\right]\right\}^{{\cal A}_j}
\end{eqnarray}
\begin{eqnarray}  
\label{ores2}
\frac{\mu_g}{M_g^o}&=&
exp\left\{\frac{16\pi}{7n-1}\left(\alpha_g^{-1}-\alpha_g^{o-1}\right)\right\}
\left[\frac{\alpha_g}{\alpha_g^o} \right]^{{\cal E}_3}
\left\{1+\frac{18\,\alpha_3^o(M_z)}{1-7n}
\left(\alpha_g^{-1}-\alpha_g^{o-1}\right)\right\}^{{\cal E}_4}
\nonumber\\
&&\times\prod_{j=1}^{3}\left\{1+\frac{7b^*_j}{1-7n}
\left[1-\frac{\alpha_g}{\alpha_g^o}\right]\right\}^{{\cal H}_j}
\end{eqnarray}
where 
\begin{equation}
{\cal E}_1={\frac{285+341n}{33(7n-1)}},\,\,\,\,\,\,\,\,
{\cal E}_2=\frac{4(3+n)}{3(1-7n)},\,\,\,\,\,\,\,\,
{\cal E}_3=\frac{2621+341n}{33(7n-1)},\,\,\,\,\,\,\,\,
{\cal E}_4=\frac{100+4n}{3(1-7n)}
\end{equation}
\begin{equation}
{\cal D}_j=\left\{\frac{(18-77n)(-4-5n)}{660(7n-1)b^*_1},
\frac{-3n(14+13n)}{4(7n-1)b^*_2},
\frac{4(2+n)(3+n)}{3(7n-1)b^*_3}\right\}_j
\end{equation}
\begin{equation}
{\cal A}_j=\left\{\frac{(-29+56n)(-4-5n)}{220(7n-1)\pi b^*_1},
\frac{81n(5+2n)}{4(7n-1)\pi b^*_2},
\frac{2(2+n)(-19+n)}{(7n-1)\pi b^*_3}\right\}_j
\end{equation}
\begin{equation}
{\cal H}_j=\left\{\frac{-268-27n+385n^2}{660(7n-1)b^*_1},
\frac{-3n(125+13n)}{4(7n-1)b^*_2},
\frac{4(50+27n+n^2)}{3(7n-1)b^*_3}\right\}_j
\end{equation}
The above analytical solution to eq.(\ref{HHMSSM})
agrees well with the  numerical one.
To find a numerical solution we just solved numerically
the system of three  equations obtained from subtracting eq.(\ref{MSSM})
from (\ref{HHMSSM}) to eliminate the $\delta_i$'s and also replaced 
$\ln(\alpha_g/\alpha_j(\mu_g))$ by
$\ln(\alpha_g/\alpha_j(\mu_g))=\ln(1+b_j^*\alpha_g/(2\pi)
\ln(M_g/\mu_g))$. The agreement between the two approaches
is good, within less than $1\%$  relative error for $\alpha_3(M_z)$,
$5\%$ relative error for the factor $M_g/M_g^o$, and $10\%$
relative error for $\mu_g/M_g^o$.
The larger error  exists when the coupling $\alpha_g$ is larger
and is also due to the presence of the logarithmic dependence the terms
involving $M_g$ and $\mu_g$ come with in the RGE equations. 

\section{Numerical Results}
In this section we analyse the results and the phenomenological
implications of eqs.(\ref{ores1}),(\ref{ores4}) and
(\ref{ores2}).
\begin{figure}[tbh]
\begin{center}
\parbox{8cm}
{\psfig{figure=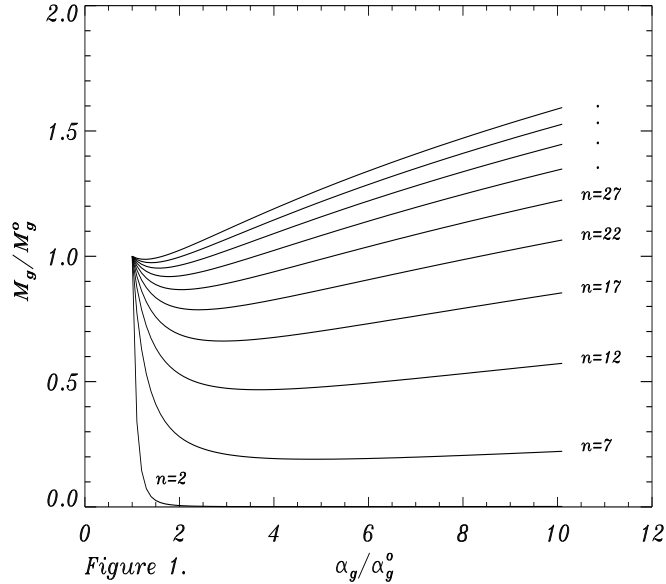,height=8
cm,width=9cm}}
\end{center}
\caption{ The values of  $M_g/M_g^o$ plotted in function of 
the ratio $\alpha_g/\alpha_g^o$ for different values of $n$.} 
\label{fig6}
\end{figure}
\begin{figure}[tbhp]
\begin{center}
\parbox{8cm}
{\psfig{figure=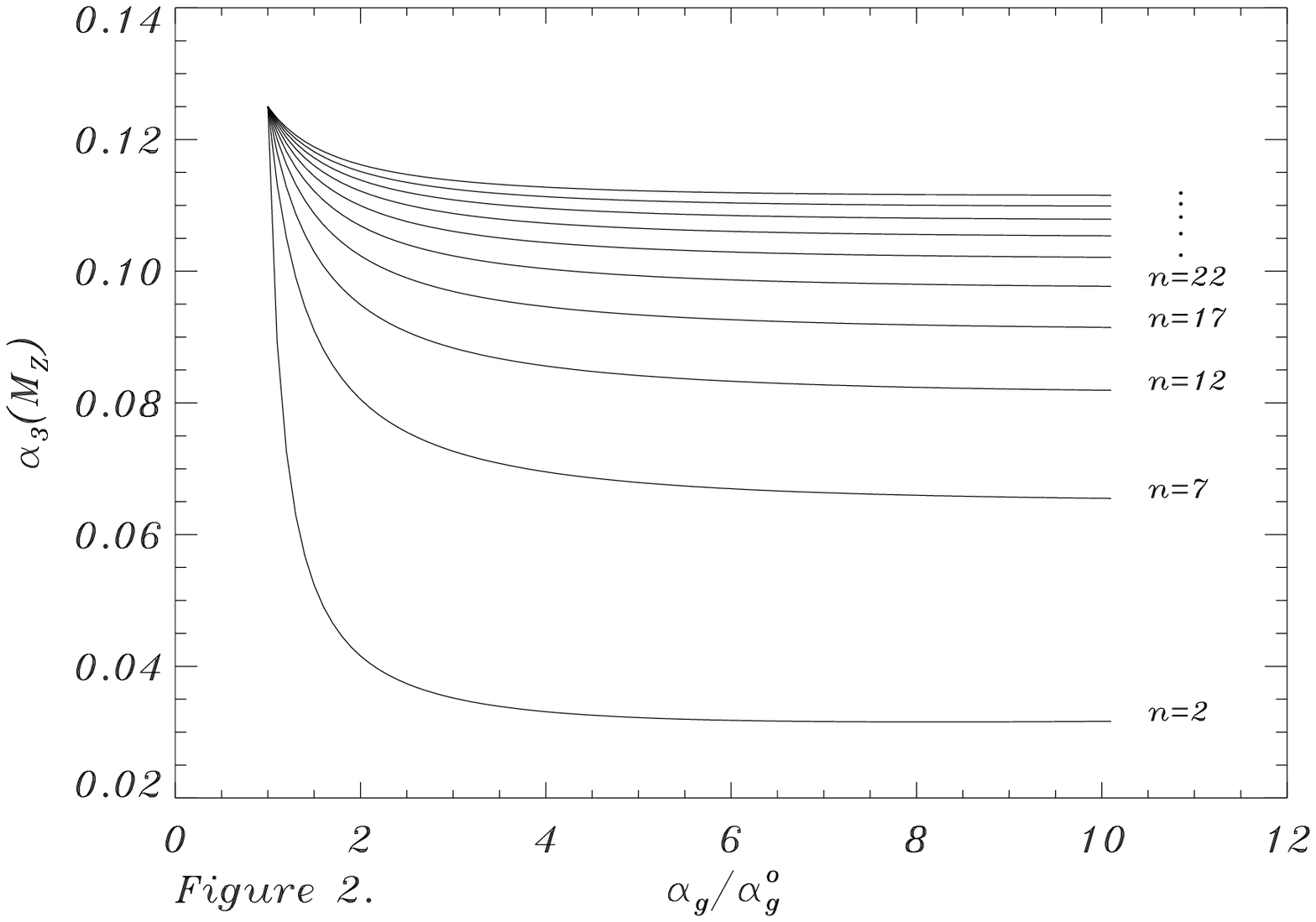,height=8cm,width=9cm}}
\end{center}
\caption{ The values of  $\alpha_3(M_z)$ plotted in function of 
the ratio $\alpha_g/\alpha_g^o$ for different values of $n$.} 
\label{fig7}
\end{figure}
\begin{figure}[tbhp]
\begin{center}
\parbox{8cm}
{\psfig{figure=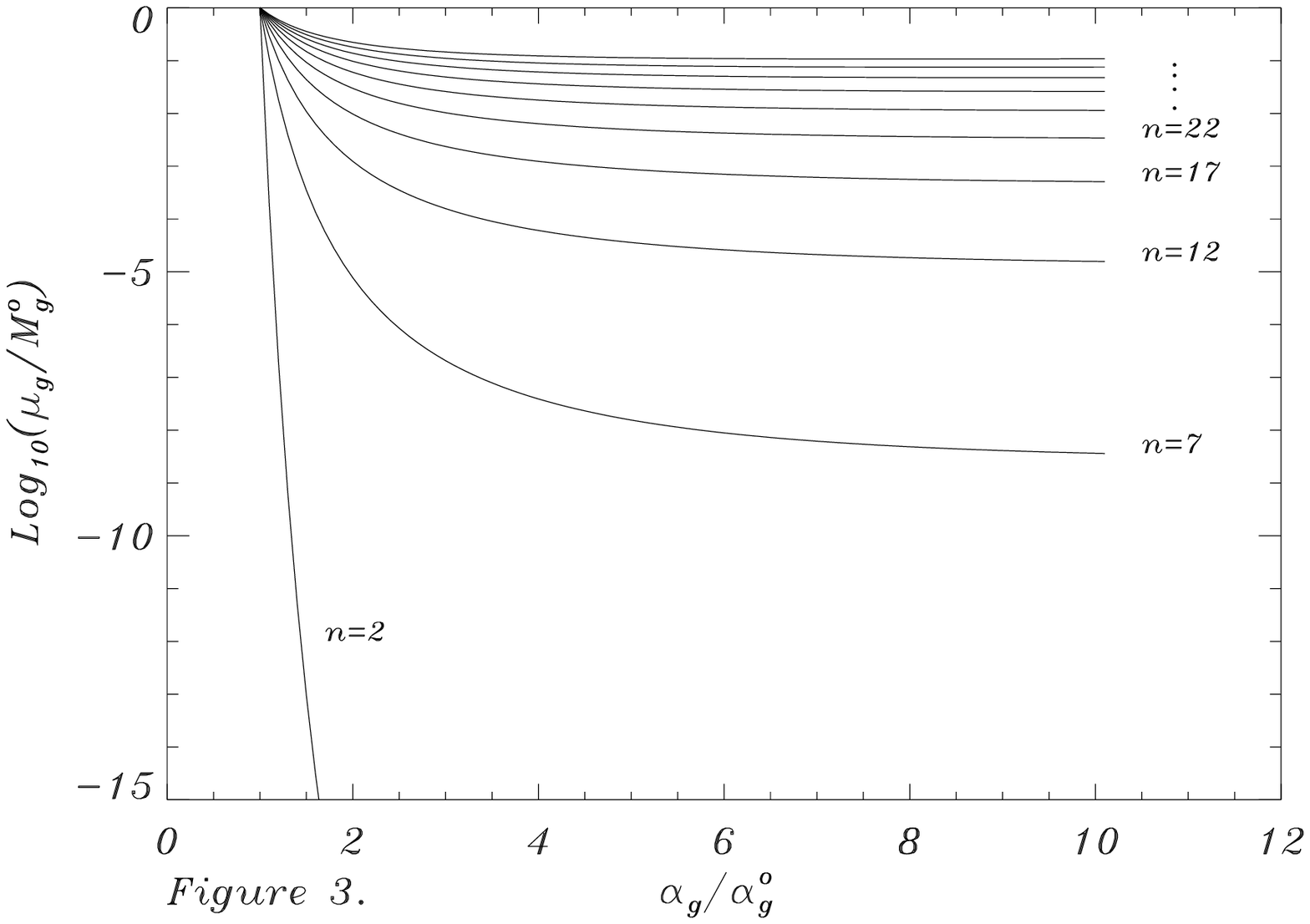,height=8
cm,width=9cm}}
\end{center}
\caption{ The values of  $Log_{10}[\mu_g/M_g^o]$ plotted in function of 
the ratio $\alpha_g/\alpha_g^o$ for different values of $n$.} 
\label{fig8}
\end{figure}
Figure 1  shows the ratio of the unification
scales $M_g/M_g^o$  for different values of $n$ in function of the ratio
$\alpha_g/\alpha_g^o$.  
We observe that this ratio is less than 
unity for most of the parameter space and the effect
of extra states we added does not bring the unification scale
closer to the weakly coupled heterotic string scale which is a factor
of $\approx 20$ above the MSSM value.
However, for {\it large} $n$
the ratio $M_g/M_g^o$
 is very close to unity, and therefore the change induced in this
case from the MSSM scale,  is very small. (Note that
the perturbative calculation is valid for large $n$, as
long as $n\alpha\approx \kappa {\cal O}(4\pi)$, with 
$\kappa <1$). For $n\geq 20$ 
we find $M_g$ above $0.8M_g^o$ for most values of $\alpha_g$,
 and therefore the change  induced by the extra matter
to the MSSM unification scale is insignificant.
This is a result of the presence of two competing effects, the
reducing of the scale (at one-loop level) 
due to the $SU(3)$ triplet states and the opposite
effect of  increasing the scale due to the  
complete five dimensional multiplets.

Such  opposite effects are also manifest in the predicted value of
$\alpha_3(M_z)$. In Figure 2 we presented this value
for different  $n$ in function of the ratio
$\alpha_g/\alpha_g^o$. We observe that we can accommodate
 values of $\alpha_3(M_z)$ smaller than in 
the MSSM and in better agreement with the experimental
value \cite{particledata} $\alpha_3(M_z)_{exp}=0.119\pm 0.002$,
provided that the value of the
unified  coupling is marginally increased from the MSSM value,
for the case of small $n$. For $n\geq 20$ $\alpha_3(M_z)$ 
is within the experimental limits for a larger range of values of $\alpha_g$.
 The effect of reducing the strong coupling 
is  essentially due to the presence of the colour triplets we
considered. The result is somewhat expected as, unlike 
models which include complete $SU(5)$ representations to the MSSM 
spectrum and 
where complete representations introduced the same
term $n\ln(M_g/\mu_g)$ in the running of the gauge couplings \cite{grl}, 
the situation here is  different because the similar contribution
is now  $(n+\Delta b_i)\ln(M_g/\mu_g)$, with $\Delta b_i$ 
standing for the triplets' contribution (see eqs.(\ref{intalph}), 
(\ref{HHMSSM})). This means that the {\it
relative} behaviour of the gauge couplings running is already changed
at one-loop order from the MSSM prediction due to the presence of the
$SU(3)$ triplets while the complete five dimensional multiplets
bring a two-loop additional increasing effect \cite{grl}.
We would  like to note
that the input MSSM value for $\alpha_3^o(M_z)$ considered here was 
$0.125$; this represents the lower limit prediction of a ``bottom-up''
approach in the  MSSM case, and therefore the predictions  we made
for $\alpha_3(M_z)$ 
could increase 
slightly if the input for $\alpha_3^o(M_z)$ is 
above this value.

Figure 3 shows  the ratio $\mu_g/M_g^o$ in terms of $\alpha_g$
and this determines the value of one of these when the other is fixed.
For the parameter space with  good predictions
for $\alpha_3(M_z)$ we find that the (bare) intermediate scale is in the
region of $3\times 10^{15}$ GeV, only a factor of $\approx 10$ below the standard
MSSM unification scale. The large value for the intermediate scale
avoids an enhancement of the proton decay rate by
the colour triplet states. 

\section{Conclusions}
In this work we have considered the phenomenological implications
of a string-motivated model,
which predicts below the compactification scale the existence
of $n$ extra  pairs $5+{\overline 5}$ of $SU(5)$ states and 2 pairs 
of $SU(3)$ triplets in addition to  the MSSM spectrum. 
The motivation for studying this model originates in the
suggestion that this might also solve the ``doublet-triplet'' splitting 
problem, commonly faced by Grand Unified Group-based theories.
The strong coupling at the electroweak
scale can be reduced below the value of the  two-loop
MSSM  prediction and be brought into better 
agreement with the experiment while the
value of the unification scale, in two-loop order, remains close
to the MSSM prediction.

\section{Acknowledgements}
The author thanks Graham Ross and Witold Pokorski for useful discussions.
D.G. acknowledges the financial support from  Oxford University 
(Leverhulme Trust research grant).

\end{document}